\documentclass[twocolumn,showpacs,preprintnumbers,amsmath,amssymb,prb,floatfix, superscriptaddress]{revtex4}

\usepackage{graphicx}
\usepackage{dcolumn}
\usepackage{bm}
\newcommand{\Cucomplex} {{\mbox{Cu${}_4$OCl${}_6$L${}_4$}}}
\newcommand{\Cudaca} {{\mbox{Cu${}_4$OCl${}_6$daca${}_4$}}}
\newcommand{\CuTeX} {{\mbox{Cu${}_2$Te${}_2$O${}_5$X${}_2$}}}
\begin{document}
\title{Tetrahedra system ${\Cudaca}$: high-temperature manifold of molecular configurations governing low-temperature properties}
\author{O. Zaharko,\cite{now}}
\author{J. Mesot,}
\affiliation{Laboratory for Neutron Scattering, ETHZ \& PSI, CH-5232 Villigen, Switzerland}
\author{L. A. Salguero,} 
\author{R. Valent\'{i},}
\affiliation{Institut f\"{u}r Theoretische Physik, Universit\"{a}t Frankfurt, D-60438 Frankfurt, Germany}
\author{M. Zbiri,} 
\author{M. Johnson,}
\affiliation{Institut Laue-Langevin, 156X, F-38042 Grenoble C\'{e}dex 9, France}
\author{Y. Filinchuk,}
\affiliation{Swiss-Norvegian Beamlines, ESRF, F-38042 Grenoble C\'{e}dex 9, France}
\author{B. Klemke,} 
\author{K. Kiefer,}
\affiliation{BENSC, Hahn-Meitner Institut, D-14109 Berlin, Germany}
\author{M. Mys'kiv,}
\affiliation{Institute for Inorganic Chemistry, Ivan Franko National University, 79005 Lviv, Ukraine}
\author{Th. Str\"{a}ssle,}
\affiliation{Laboratory for Neutron Scattering, ETHZ \& PSI, CH-5232 Villigen, Switzerland}
\author{H. Mutka}
\affiliation{Institut Laue-Langevin, 156X, F-38042 Grenoble C\'{e}dex, France}
\date{\today}

\begin{abstract}
The ${\Cudaca}$ system composed of isolated  Cu${}^{2+}$ S=1/2 tetrahedra with antiferromagnetic exchange should exhibit properties of  a frustrated quantum spin system.
$\it Ab~initio$ density functional theory calculations
for electronic structure and molecular dynamics computations suggest a complex interplay between  magnetic exchange, electron delocalization and molecular vibrations. 
Yet, extensive experimental characterization of ${\Cudaca}$ by means of  synchrotron x-ray diffraction, magnetization, specific heat and inelastic neutron scattering reveal that properties of the real material can be only partly explained by proposed theoretical models  as the low temperature properties seem to be governed by a manifold of molecular configurations coexisting at high temperatures. 
\end{abstract}

\maketitle

\section{Introduction}
Strongly frustrated quantum spin systems show, following numerous theoretical 
work\cite{Waldtmann, Moessner, Canals, Misguich, Fouet},
 highly degenerate ground states. When the degeneracy is lifted  a number of 
exotic ground states may emerge. However, in spite of large experimental
 efforts such examples are rare and are hard to probe. Any perturbation tends
 to push the real system into a conventional classical ground state. A good recent example is the ${\CuTeX}$ (X=Cl, Br) system. Being assumed to be built of weakly coupled tetrahedral units\cite{Johnsson00, Lemmens01, Brenig01, Totsuka02, Valenti03, Gros03}, it appeared to have an incommensurate  3D long-range ordered ground state\cite{Zaharko04}.\\
Our aim is to investigate the properties of tetrahedra which are 
the building blocks of geometrically frustrated quantum spin lattices (e.g. kagom\'{e},  pyrochlore).  Note that the ground state of an isolated tetrahedron with antiferromagnetic Heisenberg exchange is doubly degenerate in the case of high T${}_{d}$ symmetry. It is formed by two entangled\cite{Gros03,Bose05}
 singlets; one being a product of two dimers $\psi_1=\frac{1}{\surd4}(\mid\uparrow\uparrow\downarrow\downarrow\rangle+\mid\downarrow\downarrow\uparrow\uparrow\rangle-\mid\uparrow\downarrow\downarrow\uparrow\rangle-\mid\downarrow\uparrow\uparrow\downarrow\rangle)$ and the second composed by two-site triplet products, 
$\psi_2=\frac{1}{\surd12}(2\mid\uparrow\downarrow\uparrow\downarrow\rangle+2\mid\downarrow\uparrow\downarrow\uparrow\rangle-\mid\uparrow\uparrow\downarrow\downarrow\rangle-
\mid\uparrow\downarrow\downarrow\uparrow\rangle-\mid\downarrow\downarrow\uparrow\uparrow\rangle-\mid\downarrow\uparrow\uparrow\downarrow\rangle)$.
Three triplets and one quintet constitute the excited states.
 We choose in the present work 
 a simple frustrated quantum spin candidate among zero-dimensional systems. 

We focus our study on the ${\Cucomplex}$ system. Its main magnetic building block  is an almost ideal isolated tetrahedral cluster of four Cu${}^{2+}$ ions (see Fig.~\ref{Fig1}). The magnetic properties of several representatives of this family could not be quantitatively accounted for by a simple Heisenberg isotropic exchange\cite{Lines, Jones, Blake}.
A number of more complex models have been proposed. In early stages Lines {\it et al.}\cite{Lines} suggested the existence of orbital degeneracy of the Cu${}^{2+}$ ion and a large antisymmetric contribution to the spin Hamiltonian. However, angular overlap model calculations\cite{Chibotaru} indicated that the unpaired electron of the Cu${}^{2+}$ ion occupies the $dz^2$ nondegenerate orbital with no orbital moment, which is around 960 meV above other $d$-orbitals.\\
Henceforth the invoked models follow two routes, the first one exploits the complexity of the spin exchange and the second one puts forward the spin-vibrational interactions.
The first approach is based on the generalized spin Hamiltonian 
 \begin{eqnarray}
H_{ex}=H_{iso}+H_{as}+H_{an}=
\sum_{i,j}J_{ij}{\bf{S}}_{i}{\bf{S}}_{j}+\,
\nonumber
\\
\sum_{i,j}{\bf{G}}_{ij}({\bf{S}}_{i}\times{\bf{S}}_{j})+\sum_{i,j}\sum_{\alpha=x,y,z}J_{ij}^{\alpha}{\bf{S}}_{i}^{\alpha}{\bf{S}}_{j}^{\alpha}
\label{eq1}
\end{eqnarray}
where $H_{iso}$ is the isotropic Heisenberg-Dirac-van Vleck term, and
 $H_{as}$ and $H_{an}$  are 
 the antisymmetric and  anisotropic terms respectively. 
A number of authors\cite{Blake, Buluggiu, Black} explained magnetic susceptibility, electron paramagnetic resonance (EPR) and inelastic neutron scattering (INS) observations
by considering large antisymmetric contributions (up to 20 \% of the isotropic exchange constant). Dickinson {\it et al.}\cite{Dickinson} suggested the presence of ferromagnetic intercluster exchange in addition to the antiferromagnetic intracluster one. 

The second approach is based on the spin-vibrational Hamiltonian
 \begin{equation}
H=H_{vib}(q)+H_{ex}(S,q)
\label{eq2}
\end{equation}
where $H_{vib}(q)$ describes the harmonic vibration of the nuclei around the equilibrium position and $H_{ex}(S, q)$ depends not only on the spin S but also on the nuclear displacements $q$ of the Cu${}^{2+}$ ions. 
Jones\cite{Jones} developed a simplified model of dynamic and fluxional Jahn-Teller distortions of ${\Cucomplex}$, assuming a linear relation between the coupling constants 
and atomic shifts.  Polinger {\it et al.}\cite{Polinger1, Polinger2} extended this approach concluding that the pseudo Jahn-Teller effect is the most probable reason for the anomalous magnetic properties of the system.\\
We collected substantial experimental data on ${\Cucomplex}$ with L=diallylcyanamide=N$\equiv$C-N-(CH${}_{2}$-CH=CH${}_{2}$)${}_{2}$ (${\Cudaca}$) and performed $\it ab~initio$ density functional theory calculations for the electronic structure as well as {\it ab initio}
molecular dynamics. Our findings suggest that indeed the vibrational degrees of freedom are very important. However, the dominant ingredient is the presence of a manifold of molecular configurations
coexisting at high temperatures. Therefore both, the generalized spin and spin-vibrational Hamiltonians remain insufficient to explain the measured magnetic properties of ${\Cucomplex}$.\\

\section{Experimental details}
The ${\Cudaca}$ single crystals (Nr. 1 and Nr. 2) and the
polycrystalline sample have been prepared by reaction between CuCl and diallylcyanamide in the presence of I$_2$\cite{Gorechnick}. 
The crystal structure has been examined by X-ray synchrotron powder and single crystal diffraction performed at the BM01A Swiss-Norvegian Beamline of the European Synchrotron Radiation Facility ESRF  Grenoble, France at a wavelength of 0.7272 \AA~in the 16 K - 340 K temperature range.\\
The bulk properties have been studied using a Quantum Design PPMS commercial system. The dc- and ac- magnetic susceptibility of polycrystalline and single crystal materials have been measured in the 1.8 K - 300 K temperature range and in applied magnetic fields up to 9 T. The measurements have been performed mostly on cooling with the cooling speed of 0.5 deg/min.
The specific heat has been collected in 0.35 K -  10 K  temperature range and in applied magnetic fields up to 14.5 T, and the magnetization - in the 1.8 K - 20 K  temperature range up to 14.5 T.\\
The inelastic neutron scattering experiments have been performed on a 2 g polycrystalline protonated ${\Cudaca}$ sample on the neutron time-of-flight spectrometer FOCUS at the spallation neutron source SINQ, Paul Scherrer Institute, Switzerland and on a 0.5 g polycrystalline deuterated  sample on IN4 at the Institut Laue Langevin ILL Grenoble, France. Four setups of FOCUS have been exploited ($\lambda_i$= 5.45 \AA, 4.87 \AA,  3.42 \AA~and 2.26 \AA) to access the 0 - 13 meV energy-transfer range with various energy resolutions. On IN4 high-energy transfers up to 40 meV have been probed with $\lambda_i$=1.32 \AA, while a setup with $\lambda_i$=2.64 \AA~has been used for comparison with the FOCUS results on the protonated sample. Empty sample holders have always been measured and subtracted.\\

\section{Calculation method}
\subsection{Electronic structure}
In order to understand the electronic and magnetic properties in this compound, first principles local spin density aproximation calculations were done within the framework of the Density Functional Theory (DFT). For the description of the exchange and correlation energy, 
the Generalized Gradient Approximation (GGA) was formulated by the Perdew-Burke-Ernzerhof (PBE) density functional\cite{Perdew}.\\
In the spin-polarized version of the GGA approximation the exchange-correlation energy E$_{xc}$ is a functional of the local electron spin densities 
n$_{\uparrow}$ and n$_{\downarrow}$ and their gradients

\begin{equation}
E^{GGA}_{xc}[ n_{\uparrow}, n_{\downarrow}]=\int f( n_{\uparrow}, n_{\downarrow},  \nabla n_{\uparrow}, \nabla n_{\downarrow}).
\end{equation}

For calculating the eigenvalues and eigenvectors in the DFT scheme the Full Potential Linearized Augmented Plane Waves method (FP-LAPW), as implemented in the Wien2k code \cite{wien2k}, has been used. In this method the unit cell is divided in two regions, the first one is composed by non-overlapping spheres centered at the atomic sites and the second is the region between the atomic spheres or interstitial region. In the first region a spherical potential is assumed and therefore the wave functions are expanded in terms of spherical harmonics. In the interstitial region a constant potential is assumed. This allows to expand the wave functions in that region in terms of plane waves.\\
In our calculation the valence wave functions have been expanded up to $l$=10. Sphere radii R$_{mt}$ of 2.1 a.u. for the Cu atom, and  1.6, 1.4 , 1.2, 0.9 and 0.84 a.u. for Cl, O, N, C and H atoms respectively have been set. Within the Cu muffin-tin radii chosen we excluded the 4$s$ states (the average radii for the wave functions belonging to these states is 2.7 a.u.), therefore these states lie in the interstitial region. The Cu-3$d$ and 3$p$ states enter in the calculation as the valence and semicore states respectively. For the Cu 3$d$ states the LAPW basis with augmented plane wave plus local orbital [APW+LO] method have been used. The plane-wave expansion with R$_{MT}$K$_{MAX}$ equal to 3.76 has been set.  This is a typical value when C-H bonds are present.  For the integration in the irreducible wedge,  48 k-points have been used.

\subsection{Molecular dynamics}
The dynamics of Cu-daca has been investigated using the DFT code VASP
\cite{Kresse1, Kresse2} which is well adapted to a system as large as ${\Cudaca}$ (174
atoms per unit cell). VASP uses a plane wave basis set and the Projector
Augmented Wave (PAW)
pseudo-potentials to describe the core electrons of the atoms. The
GGA-PBE functional was used\cite{Perdew} and all electronic calculations were
performed at the $\Gamma$ point in reciprocal space.
Since phonons and molecular vibrations are of interest in this paper,
lattice dynamics calculations, involving the determination of the
dynamical matrix, were attempted. In view of the number of atoms in the
unit cell, the direct method \cite{Parlinski} is the only practically
feasible approach in which symmetry-inequivalent atoms are displaced one
by one from equilibrium and the Hessian is constructed from the
Hellmann-Feynman forces from the series of electronic calculations. \\
In the case of ${\Cudaca}$, the direct method always resulted in negative frequencies. This method invokes a significant perturbation to 
the equilibrium configuration and therefore the electronic structure.
Indeed, electronic and structural correlations between the core and the ligands lead to an electronic structure which is very sensitive 
to the precise molecular configuration. Such sensitivity is an indication of the fragility of this molecular system as proved
by the physical meaning of negative frequencies in this case.\\
In this situation we resorted to $\it ab~initio$ molecular dynamics, using
VASP, in which case the (partial) vibrational density of states
((p)VDOS) can be obtained from the velocity auto-correlation function
and the scattering function is obtained from the van Hove correlation
functions. Both of these correlation functions are determined directly
from the $\it ab~initio$ molecular dynamics (AIMD) trajectories using the nMoldyn program \cite{nMoldyn}. The
temperature dependence of the dynamics, which may reveal
anharmonicities, is investigated via the average kinetic energy of the
atoms. The drawback of the molecular dynamics approach compared to a lattice dynamics
calculation, is that the phase relation of the atomic displacements is
lost and normal modes and their frequencies cannot be determined. One
solution is to investigate the time-dependent fluctuations of geometric
quantities associated with groups of atoms. In the case of the magnetic
properties of ${\Cudaca}$ the Cu tetrahedron is of interest.\\
AIMD simulations have therefore been performed at 100 K and 250 K for a
protonated crystal. Production simulations at constant volume and energy (NVE ensemble) cover ~10 ps, that is $\approx$10$^4$ simulation steps. The 100 K simulation was then re-equilibrated for
a deuterated crystal and the NVE production run in this case lasted 27
ps in order to give better access to the low frequency modes.\\

\section{Results}

\subsection{Crystal structure}
\begin{figure}[tbh]
\caption {(Color online)The $ab$ projection of the ${\Cudaca}$ unit cell. The Cu atoms are represented by red circles, Cl, N and C - by green, blue and grey circles, respectively. H atoms are omitted for clarity, O atoms are located behind Cl2 and, therefore, are not visible.}
\includegraphics[width=86mm,keepaspectratio=true]{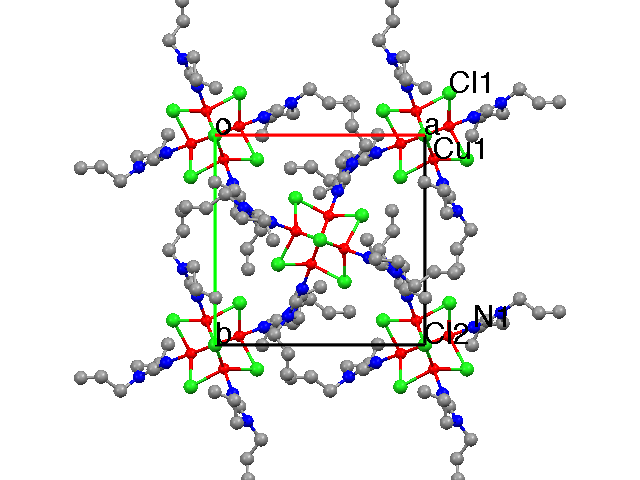}
\label{Fig1}
\end{figure}

\begin{figure}[tbh]
\caption {Difference Fourier map of ${\Cudaca}$ at 80 K (top) and 300 K (bottom) from x-ray single crystal data.}
\includegraphics[width=86mm,keepaspectratio=true]{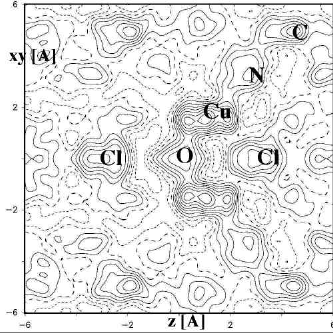}
\includegraphics[width=86mm,keepaspectratio=true]{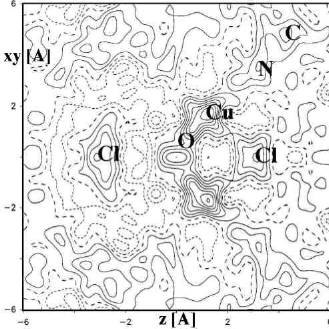}
\label{Fig2}
\end{figure}

The ${\Cudaca}$ structure is tetragonal\cite{Gorechnick} (space group $P\overline{4}2_{1}c$) in the whole 16 K - 340 K studied temperature interval.
There are two ${\Cudaca}$ molecules per unit cell and they are slightly twisted around the $z$ axis (Fig.~\ref{Fig1}). The oxygen is in the center of the molecule, surrounded by a tetrahedron of Cu${}^{2+}$ ions with two different Cu-Cu distances, equal to 3.0790(2) \AA~and 3.1185(2) \AA~at 80 K. The tetrahedron is enclosed by a distorted octahedron of chlorine ions with four Cl$_{1}$ in the square base and two Cl$_{2}$ in the apexes. Each copper has trigonal bipyramidal coordination: three  Cl${}^{-}$ ions  form the triangular base of the bipyramid, while the central O${}^{2-}$ ion and a terminal ligand L are located in apical positions. 
The exchange paths Cu-O-Cu are identical, while the two paths through chlorine ions, Cu-Cl$_{1}$-Cu and Cu-Cl$_{2}$-Cu,  slightly differ (see Table~\ref{tab1}). Such geometry implies rather high symmetry of the cluster and of the spin exchange Hamiltonian, only marginaly deviating from T${}_{d}$.\\
There is an anomaly in the lattice constants in the 230 K - 280 K temperature range notifying an isostructural order-disorder crossover\cite{note1}. Apparently the molecule attains several configurations with small energy 
difference between them. Above T$_{C}$=282 K these configurations coexist, while below T$_{C}$ molecules freeze in one of them. The configurations differ mainly by the position of the branched diallylcyanamide ligands, which fold. This modifies the distances between the molecules (Table~\ref{tab1}) and the unit cell volume shrinks by 7 \% at 80 K compared to the volume at room temperature. The Cu$_{4}$OCl$_{6}$N$_{4}$ core changes very little. The only noticeable difference is the shift of the Cl$_{1}$ ions by 0.2487(9) \AA, while for Cu and Cl$_{2}$ the shift is only 0.0462(9) \AA~and 0.0282(9) \AA, respectively.\\
Inspection of the interatomic distances and angles presented in Table~\ref{tab1} suggests that the Cu-O-Cu  magnetic exchange path is antiferromagnetic (AF), while the Cu-Cl-Cu one is ferromagnetic (F). Therefore the mobility of the Cl$_{1}$ ions is essential in determining the magnetic exchange and this is investigated with the help of AIMD simulations.\\
One more appreciable detail could be extracted from the single crystal diffraction data. The difference Fourier map through the N-O-N plane of the molecule (Fig.~\ref{Fig2}) shows additional electron density near Cu atoms at 80 K. The tiny peaks (0.38 and 0.31 e/\AA$^3$) are displaced along $z$ by 0.84(5) \AA~and 0.73(10) \AA~from Cu towards O and N (Figure~\ref{Fig2}), respectively. At 300 K these peaks are less pronounced (0.24 e/\AA$^3$). These difference electron density peaks near Cu and O, N, C positions evidence the delocalization of e-density in the O-Cu-N$\equiv$C bonds. No smearing of the electron density near copper atoms or an increase of their anisotropic displacement parameters is observed;
the mean-square displacement amplitudes are very small (see Table~\ref{tab1}). This implies the absence of a static and/or dynamic Jahn-Teller effect at low temperatures.
\begin{table}
\caption{Comparison of the characteristic angles [deg], intra- (d$_{intra}$ [\AA]) and inter- (d$_{inter}$) molecule distances in ${\Cudaca}$ at 80 K and 300 K.
Mean-square displacement amplitudes of Cu (MSD$_{Cu}$, [\AA$^2$]) and other atoms (MSD$_{lig}$, [\AA$^2$]) in the specified bond directions of the CuCl$_3$ON trigonal bipyramid calculated from the anisotropic displacement parameters refined at 80 K.}
\label{tab1}
\begin{ruledtabular}
\begin{tabular}{llllll}
d$_{intra}$& 80 K&300 K&  Angles&80 K&300 K\\
Cu-Cu & 3.0790(2)& 3.0933(4)&$\angle_{CuOCu}$&108.106(6)& 108.9(2)\\
Cu-Cu & 3.1183(2)& 3.1101(5)&$\angle_{CuOCu}$&110.158(6)& 109.8(2)\\
Cu-O   & 1.9016(2)& 1.9011(3)&$\angle_{NCuO}$&179.01(4)& 178.8(1)\\
Cu-N1 &1.932(1)& 1.935(3)&$\angle_{CuCl1Cu}$&80.46(1)& 80.48(3)\\
Cu-Cl1& 2.3694(4)& 2.3911(9)&$\angle_{CuCl2Cu}$&80.59(1)& 80.62(3)\\
Cu-Cl2& 2.3806(3)& 2.3909(8)&$\angle_{Cl1CuCl1}$&117.97(2)& 118.85(4)\\
Cu-Cl1& 2.4578(3)& 2.4231(9)&$\angle_{Cl1CuCl2}$&127.73(1)& 120.92(3)\\
\\
&MSD$_{Cu}$&MSD$_{lig}$&d$_{inter}$&80 K&300 K\\
Cu-O   &0.0107 & 0.0125 &d$_{100}$&12.60490(3)&12.5652(1)\\
Cu-Cl1& 0.0143 & 0.0130&d$_{001}$&12.73045(5)&13.7685(2)\\
Cu-Cl2& 0.0203 & 0.0189&d$_{111}$&10.9525(3)&11.2399(1)\\
Cu-N1&  0.0108 & 0.0152&&&\\
\end{tabular}
\end{ruledtabular}
\end{table}
\subsection{Electronic structure and charge density distribution}
\begin{figure}[tbh]
\caption {(Color online) Partial Density of States in the non-spin polarized GGA calculations for ${\Cudaca}$ at T=80 K.}
\includegraphics[width=86mm,keepaspectratio=true]{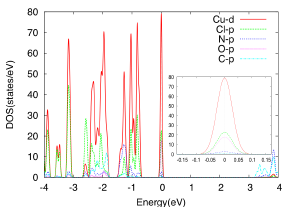}
\label{Fig1DFT}
\end{figure}

The electronic  Density of States (DOS) has been calculated for
 the crystal structures determined at temperatures, 80 K and 340 K.
Since both structures have almost indistinguishable DOS features, we present here the
results for the T=80 K structure.
In Figure ~\ref{Fig1DFT} we show the partial Cu $d$, Cl $p$, N $p$, O $p$ and C $p$ DOS in the
range of energies between -4 and 4 eV obtained from non-spin polarized
 GGA\cite{Perdew} calculations.
We observe a very narrow and sharp contribution of Cu $d$ and Cl, N and O $p$
 states at the Fermi level
well separated from the rest of the valence and conduction states.
 These features  indicate that these
states strongly hybridize and that the system is almost zero
 dimensional, formed by well
isolated Cu tetrahedra. The bandwidth at the Fermi level is less than 0.2 eV (see the inset of Fig.
~\ref{Fig1DFT} where a blow up of the DOS at the Fermi level is shown). From calculations
of the crystal field splitting we conclude that the Cu $d$ band at the Fermi level has
$d_{z^2}$ character (in the local reference frame of the CuCl$_3$ON trigonal bipyramid).

\begin{figure}[tbh]
\caption {Partial Density of States for majority (upper panel) and minority (lower panel) spin contributions from Cu-$d$ states of ${\Cudaca}$ at T=80 K.}
\includegraphics[height=86mm,keepaspectratio=true,angle=-90]{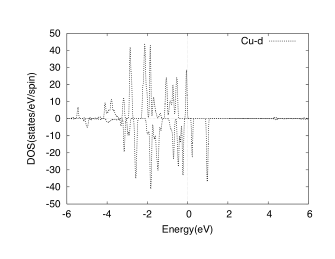}
\label{Fig2DFT}
\end{figure}

\begin{figure}[tbh]
\caption{(Color online) Charge density distribution at T=80 K.}
\includegraphics[width=86mm,keepaspectratio=true]{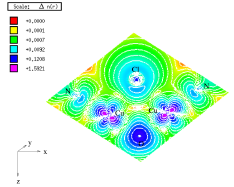}
\label{Fig3DFT}
\end{figure}

Within the GGA approximation\cite{Perdew} we obtain that the Cu $d$ states are partially occupied and
therefore the system is metallic. This is a well-known drawback of DFT calculations within the LDA
or GGA approach when performed for correlated systems. Improved exchange-correlation functionals
like LDA+U \cite{Anisivom91} should provide
the correct insulating behavior of the material. In the case of
${\Cudaca}$, calculations
at the level of spin-polarized GGA
 already open a small gap at the Fermi level as shown in Fig. \ref{Fig2DFT}. In this figure
we present the partial DOS for Cu $d$ states in both spin up (upper pannel)
 and spin down (lower pannel) channels. The occupation of the spin-up and spin-down Cu $d$ states is very similar,
 therefore the resulting localized magnetic $d$-moment  is small.
Table~\ref{tab3} presents the magnetic moment at the cluster core inside the muffin-tin radii.
 The unpaired Cu electron  is strongly delocalized via hybridization mainly with oxygen
 and less strongly with chlorine and nitrogen.\\
In Fig.~\ref{Fig3DFT} we show the projection on the $xy$ plane of the
  calculated charge density map of the band around the Fermi level.
We note the good agreement  with the measured
difference Fourier map from the 80 K x-ray diffraction experiment (Fig. \ref{Fig2}).

\begin{table}
\caption{Magnetic moment (M[m$_{B}$/atom]) within the muffin-tin radii ($r_{mt}$) and its total moment fraction [\%]. For the interstitial the moment is normalized per formula unit 
(f.u.=CuO${}_{0.25}$Cl${}_{1.5}$daca).}
\label{tab3}
\begin{ruledtabular}
\begin{tabular}{cccc}
Atom&r$_{mt}$ &  M &  \\ 
Cu& 2.1 & 0.58350& 58\\
O& 1.4 &0.44204 & 11\\
Cl1& 1.6 & 0.08352&8.3\\
Cl2& 1.6 & 0.09027&4.5 \\
N1&  1.2 & 0.02857 &2.9\\
interstitial& &0.14477 &14\\ 
\end{tabular}
\end{ruledtabular}
\end{table}
\subsection{Bulk properties}
\begin{figure}[tbh]
\caption {(Color online) Temperature dependence of dc-magnetization (M/H) of the single crystal Nr. 1 measured at 0.1 T, 0.2 T, 0.5 T, 4 T and  of polycrystalline (black solid line) ${\Cudaca}$
measured at 0.1 T (left scale). (M/H)$\cdot$T of crystal and powder measured at 0.1 T (right scale). Inset: zoomed 1.8 K - 30 K temperature interval of M/H of the crystal.}
\includegraphics[width=86mm,keepaspectratio=true]{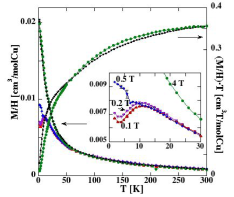}
\label{Fig3a}
\end{figure}

The temperature dependence of dc-magnetization (M/H) measured for the
polycrystalline sample at H=0.1 T and for a single crystal (Nr. 1) at several values of applied magnetic field is presented in Fig.~\ref{Fig3a}. At 300 K the (M/H)$\cdot$T value is 0.375 cm$^3$(molCu)$^{-1}$K suggesting that the system is composed of uncoupled S=1/2 spins. 
The Curie-Weiss temperature estimated from the high-temperature susceptibility is $\theta_1$=$-$50 K, which implies dominant antiferromagnetic interactions.\\
For the polycrystalline sample M/H gradually increases with temperature lowering and at 1.8 K reaches 0.02 cm$^3$(molCu)$^{-1}$. This is a very small value, suggesting the presence of a tiny amount of free S=1/2 impurities.  However, the temperature dependence cannot be fitted to a Curie-like 1/T contribution, which would grow much faster at low temperatures.\\
For a single crystal,  dc-magnetization consists of two major contributions. The first contribution has a bump near 12 K and can be attributed to isolated tetrahedra with antiferromagnetic exchange with J=-1.8 meV resulting in a spin-singlet ground state. 
The second contribution increases with temperature decrease and, similarly to the polycrystalline case, cannot be fitted to the paramagnetic 1/T term. 
The ratio of these contributions is different for different single crystals and depends on crystal orientation and applied magnetic field. 
Based on a number of dc-magnetization measurements we reckon that the ratio of these contributions and the sensitivity to magnetic field
is related not to a specific crystallographic direction but to the morphology of crystal. Apparently there is a strong growth anisotropy and not a
lattice anisotropy.\\
\begin{figure}[tbh]
\caption {(Color online) Temperature dependence of dc-magnetization (M/H) of the single crystal Nr. 2 measured with magnetic field H=0.2 T applied 
along three crystallographic directions: [110] (green circles), [001] (red triangles) and [1-10] (blue squares), corresponding to short, middle and long edges of the crystal.
Inset: Field dependence of dc-magnetization of single crystal Nr. 2 and powder ${\Cudaca}$ at 1.8 K. }
\includegraphics[width=86mm,keepaspectratio=true]{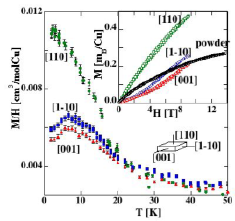}
\label{Fig4}
\end{figure}
Crystals usually grow as platelets with the [110] axis directed along the shortest edge, and [001] and [1-10] being in the basal plane.
An example of M/H utmost sensitive to the field is presented in inset of Fig.~\ref{Fig3a}. The measurement is performed with magnetic field applied along the shortest edge. 
With increasing field the bump at low temperatures decreases and already at H=0.5 T vanishes.\\
For another crystal (Nr. 2) M/H shows no bump along the shortest edge even at H=0.2 T (Fig.~\ref{Fig4}), though when the magnetic field is applied along the middle [001] or long [1-10] edges of the crystal, the bump is tractable to much higher fields (not shown).\\
One more important observation is that the low-field (0.05 T) dc-magnetization curves measured for the zero field cooled and 0.05 T cooled conditions are different (not shown). This implies that cooling in a relatively small field is sufficient to change the microscopic state of the crystal.\\
The field dependence of dc-magnetization is also sample and direction
 dependent (Fig.~\ref{Fig4} inset). For the polycrystalline sample magnetization is concave and does not saturate at 14 T reaching only 0.27 $m_{B}$/Cu. For the single crystal Nr. 2 magnetization attains almost 0.5 $m_{B}$/Cu for the [110] direction, while for other directions it approaches values similar to the polycrystalline sample.\\
The most plausible origin of our observations is the presence of
nonuniform intratetrahedral exchange couplings within a single crystal and - even to larger extent - within a polycrystalline sample. This diversity most probably appears due to the manifold of molecular configurations above the order-disorder crossover at T$_C$=282 K, which could influence the microscopic state of the system in two ways: i) during the crystal growth, performed at room temperature, several non-ground state molecular configurations freeze-in, ii) during the field cooling through the crossover region the applied field favors certain non-ground state molecular configurations.
In these frozen-in molecule configurations the Cl$_1$ ions could have positions which strengthen the ferromagnetic exchange path between two or more Cu${}^{2+}$ ions. As these molecular configurations are minority and do not represent the ground state, we shall call them 'defects' for short.\\
In agreement with the above considerations it is reasonable to assume a coexistence of at least two kinds of clusters to explain the measured dc-magnetization: i) AF ones, with antiferromagnetic (AF) exchange of the order of J=-1.8 meV, ii) AF/F ones, with ferromagnetic (F) couplings of the similar strength as the AF ones. The susceptibility calculated for this case (Fig.~\ref{Fig5} 
red curve) fits relatively well the experiment. We can fit our data also with the model comprising large anisotropy of exchange within the single type of clusters (Fig.~\ref{Fig5} blue curve), but  an unreasonably large anisotropy parameter J$_z$=1.2 meV is required. We therefore give preference to the hypothesis of the presence of two kinds of clusters.\\
\begin{figure}[tbh]
\caption {(Color online) Experimental dc-magnetization (M/H) of the single crystal Nr. 1(black) and the calculated susceptibility for
i) two types of clusters, AF and AF/F
with two AF J$_{1}$=-1.8 meV, four F J$_2$=1.8 meV (red), ii) one type of tetrahedral clusters with J$_{iso}$=-1.8 meV, J$_z$=1.2 meV (blue).}
\includegraphics[width=86mm,keepaspectratio=true]{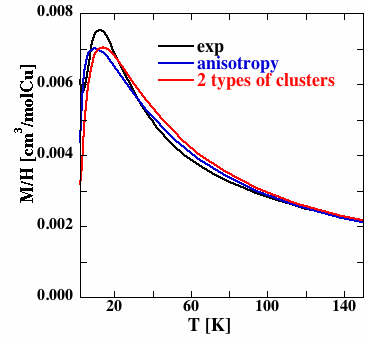}
\label{Fig5}
\end{figure}
Within the proposed scenario it is also easy to understand the strong sensitivity of the bulk properties to the applied magnetic field which is significantly smaller than $J$. The real part of the ac-susceptibility of polycrystalline sample shows several steps as a function of applied field (Fig.~\ref{Fig6}). The most pronounced step is between 2 T - 4 T, which corresponds to the Zeeman energy of 0.24 meV - 0.48 meV. 
Also the low temperature specific heat of polycrystalline and single crystal samples contains a broad Schottky-type anomaly centered at 0.4 K (0.034 meV) in 0-field data (Fig.~\ref{Fig7}). This feature shifts to higher temperatures with an applied field. 
The magnetic entropy extracted below 10 K after subtraction of a constant phonon contribution reaches only 0.7-0.8 J/KmolCu. This is approximately 1/8 of the Rln2=5.763 J/KmolCu value expected for the spin degree of freedom, so this magnetic contribution corresponds to a minor part of the sample. We deduce that one of the triplet E-levels of the AF/F clusters is situated at 0.4 K, though collected data are insufficient to propose the full energy scheme for AF/F clusters.
\begin{figure}[tbh]
\caption {(Color online) Real part of ac-susceptibility of polycrystalline ${\Cudaca}$ measured in an applied magnetic field up to 9 T at 1.8 K, 3 K and 6 K.}
\includegraphics[width=86mm,keepaspectratio=true]{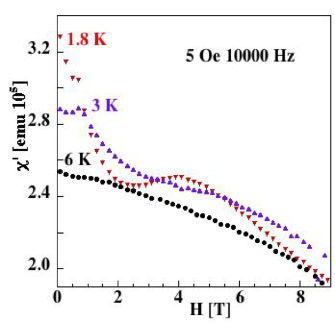}
\label{Fig6}
\end{figure}
\begin{figure}[tbh]
\caption {(Color online) Log/log plot of specific heat of polycrystalline (top) and single crystal (bottom) ${\Cudaca}$ in the 0.28 K - 10 K temperature range with applied magnetic field 0 T - 12 T.}
\includegraphics[width=86mm,keepaspectratio=true]{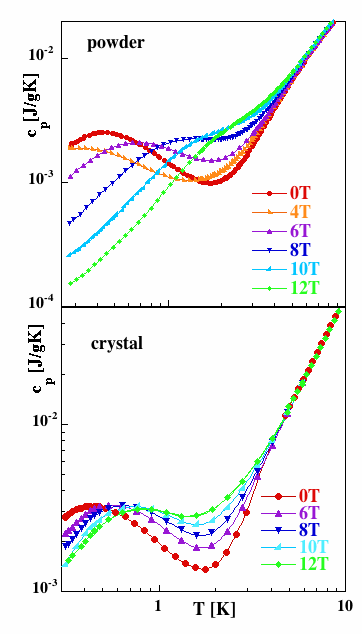}
\label{Fig7}
\end{figure}
\subsection{Inelastic neutron scattering}
\begin{figure}[tbh]
\caption {(Color online) INS spectra of protonated ${\Cudaca}$ at T=2 K (blue) and 35 K (red) measured on FOCUS with $\lambda_i$=4.85 \AA. 
The intensity has been integrated over the momentum transfer range 0.5 $<$ Q $<$ 2 \AA$^{-1}$.}
\includegraphics[width=86mm,keepaspectratio=true]{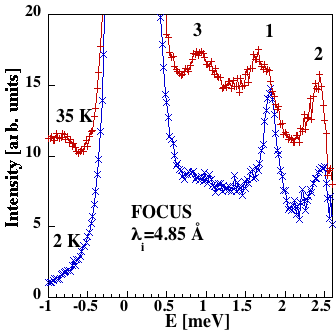}
\label{Fig1INS}
\end{figure}

\begin{figure}[tbh]
\caption {(Color online) INS spectra of deuterated ${\Cudaca}$ at T=2 K (top) and 75 K (bottom) measured on IN4 with $\lambda_i$=2.64 \AA~for three Q-values 1.5 (blue), 2.5 (violete) and 3.5 (red) \AA$^{-1}$, with increasing intensity  integrated over 0.5 \AA$^{-1}$.}
\includegraphics[width=86mm,keepaspectratio=true]{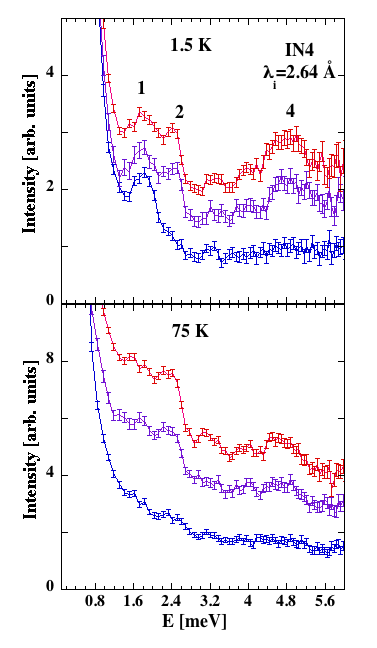}
\label{Fig3INS}
\end{figure}

\begin{figure}[tbh]
\caption {(Color online) INS spectra of deuterated ${\Cudaca}$ at T=2 K and 75 K measured on IN4 with $\lambda_i$=1.32 \AA.
The intensity has been integrated for two Q-values 3.5 and 5.5 \AA$^{-1}$ over 1 \AA$^{-1}$.}
\includegraphics[width=86mm,keepaspectratio=true]{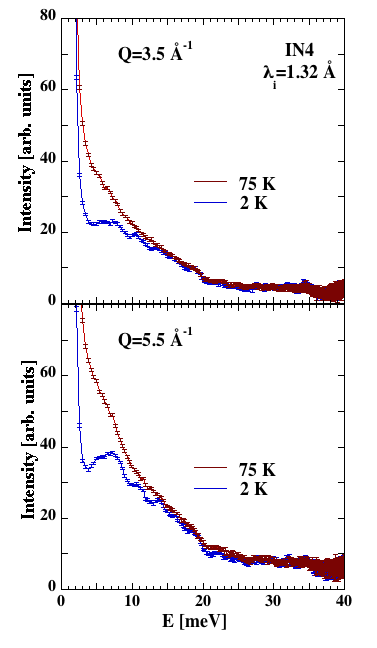}
\label{Fig5INS}
\end{figure}

For the protonated sample at 2 K two sharp features have been detected  at 1.8 meV and 2.4 meV on the neutron energy loss side, they are labeled 1 and 2 in Fig.~\ref{Fig1INS}. The peaks are at least two times broader than the resolution function (0.08 meV at 1.8 meV). At higher temperatures (7 K, 15 K, 35 K) 
the 1.8 meV feature broadens and shifts to lower energies or a 2nd peak at slightly lower energy of 1.6 meV appears and increases in intensity, while the 1.8 meV peak decreases. The 2.5 meV feature increases in intensity and an additional feature at 1 meV (labeled 3) appears. No extra peaks have been observed in the low energy region down to 0.4 meV.\\
A number of peaks are observed at higher energy transfers (not shown in Fig.~\ref{Fig1INS}), of which the most pronounced is the one at 4.8 meV. Its width is at least three times broader 
than the resolution function (0.3 meV at 4.8 meV) and intensity increases from 2 K to 35 K.\\
The measurements on the deuterated sample reveal essentially the same excitation spectrum (Fig.~\ref{Fig3INS}). The 1.8 meV peak tends to vanish at 75 K, while the 2.4 meV and 4.8 meV features decrease from 2 K to 75 K. From the Q(momentum)- and T(temperature)-dependence we relate the 1.8 meV peak to a spin singlet-triplet transition between the cluster levels, though it is not clear if it is a single excitation or several ones. It clearly corresponds to the bump in M/H. The 4.8 meV feature is increasing with Q and most probably is related to molecular vibrations, while for the 2.4 meV and other features the FOCUS and IN4 results are less conclusive.\\   
The measurement at higher energy transfers showed that relatively sharp peaks discussed above reside on a very broad feature (up to 20 meV), which increases in intensity with temperature and Q (Fig.~\ref{Fig5INS}). These spectral features and their magnetic or phonon origin are understood with the help of AIMD simulations (below), which reproduce the inelastic scattering response and allow deeper insight via partial vibrational densities of states and specific two, three or four atom correlation functions.
\subsection{Molecular dynamics}

\begin{figure}[tbh]
\caption{(Color online) Calculated coherent scattering function (CSF) for four Q-values ranging, with increasing intensity, from 2.5 to 5.5 \AA$^{-1}$ of deuterated ${\Cudaca}$.}
\includegraphics[width=80mm,keepaspectratio=true]{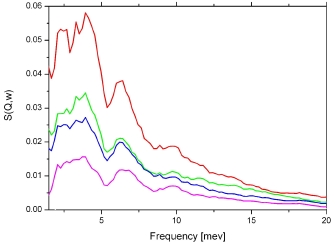}
\label{Fig2MD}
\end{figure}

\begin{figure}[tbh]
\caption{(Color online) Vibrational density of states for each atom type (top). The vertical, dashed line highlights pronounced minima at 5 meV in all partial DOS
(except for the oxygen vDOS) and the existence of lower frequency modes. Examples of latter for deuterium, carbon and copper, are shown separately (down).}
\includegraphics[width=80mm,keepaspectratio=true]{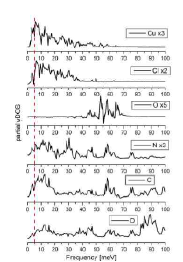}
\includegraphics[width=80mm,keepaspectratio=true]{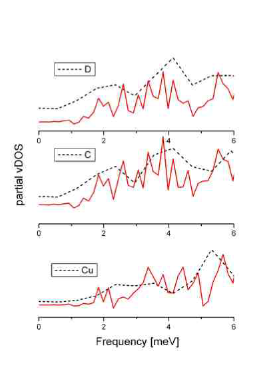}
\label{Fig3MD}
\end{figure}

\begin{figure}[tbh]
\caption{Two-atom (bond) and four-atom (twist angle) correlations as a function of frequency from the MD calculation.}
\includegraphics[width=80mm,keepaspectratio=true]{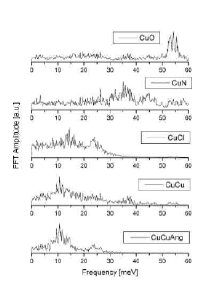}
\label{Fig4MD}
\end{figure}

AIMD simulations give a direct view of the amplitude of atomic
displacements. 
While the light atoms tend to have larger displacements than the heavier
atoms (Cl and Cu), this is not the case for O which is trapped in the
Cu-tetrahedron and the Cl-octahedron. Note that the atoms are classical particles in the AIMD simulations so that their displacements do not include zero-point motion, which can be of the order of 30\% for hydrogen at room temperature. Calculated mean-square displacements (MSD) averaged over all atoms of the same type are 0.0177 \AA$^2$ for Cu, 0.0300  \AA$^2$ for Cl, 0.0093 \AA$^2$ for O and 0.0283 \AA$^2$ for N. This is in good agreement with values obtained by projection on specified bond directions from the 80 K crystal structure refinement listed in Table~\ref{tab1}.\\
The MD simulations are validated by comparing calculated and measured
scattering functions. We note that the low frequency modes
are independent of hydrogen or deuterium (H/D) substitution and must therefore be
delocalized modes involving many atoms.
 An acoustic phonon would show a 2\% frequency shift upon deuteration. Fig.\ref{Fig2MD} shows the coherent scattering function 
$S_{coh}(Q,\omega)$ from the
27 ps simulation on the deuterated system.  Spectra at lower Q-values have weak intensity and are not reliably calculated with our model. Low frequency, numerical noise in the vibrational Density of States
(vDOS) 
contributes strongly to the coherent structure factor since this quantity is proportional to vDOS/frequency. At high Q there are peaks at 2 meV, 4.5 meV and 7 meV and a shoulder at  $\approx$10 meV which are consistent with the experimental data (Figs.~\ref{Fig1INS}-\ref{Fig5INS}). Since the AIMD simulations do not include spin degrees of freedom, these peaks can only be the signature of lattice excitations and, as expected, their intensities increase with increasing Q. The coherent scattering function is only calculated at the reciprocal lattice points of the simulation cell so that Q-sampling is limited to the Brillouin zone centre when using a single cell for the MD simulations. Simulated spectral peaks therefore correspond to $\Gamma$-point modes.\\
Atomic contributions to the signal are obtained from the 
partial vDOS (pvDOS)
which are shown in Fig.~\ref{Fig3MD} top. Each pvDOS has been normalized to the
corresponding number of degrees of freedom, 3$n$, where $n$ is the number of
a particular species. Fairly well-defined peaks exist at the lowest
frequencies in the low temperature MD simulations. The delocalized
nature of the modes is shown by the fact that all atoms have similar
pvDOS's below $\approx$5 meV. 
This result is consistent with there being negligible spectral shift upon deuteration. The trapped oxygen atom has a characteristic vibration frequency of the order of 55 meV and this is the only atom that does not contribute to the lowest frequency modes. The analysis of the chlorine atoms has been refined to separate the contributions from apical atoms and those in the square base around the copper tetrahedron. There is a small but significant softening of the apical-chlorine vibrations compared to those of the square-base chlorines, average frequencies and standard deviations being, respectively, 18.8$\pm$13.5 meV and 20.3$\pm$14.4 meV. Fig.~\ref{Fig3MD} bottom
shows examples of  the lowest frequency part of the vDOS for D, C and Cu for different lengths of simulations. The dotted curves show the results from a 10 ps trajectory while the solid curves are derived from the 27 ps trajectory. The longer trajectory gives better frequency resolution which shows that the first peaks in the DOS occur at 2 meV. Spectral peaks below this frequency must therefore be of magnetic origin.\\ 
Features of the VDOS can be further understood by considering n-atom correlations. The frequency spectrum of interatomic distances involving Cu atoms is shown in Fig.\ref{Fig4MD}. The Cu-Cu stretch is strongest at $\approx$12 meV. The twisting deformation of the tetrahedron, that is the angle variation between Cu-Cu vectors on opposite sides of the tetrahedron, also has a characteristic frequency of $\approx$12 meV. These are the modes that could directly modulate the magnetic copper-copper interactions. 
The oxygen pathway is clearly not modified by any low frequency modes whereas the copper-chlorine distance modulations show a significant amplitude from 25 meV down to the lowest frequencies.\\
Spin restricted AIMD simulations are the only way to gain insight into
the low frequency modes measured on FOCUS and IN4.
Simulations show excitations at above 2 meV and at $\approx$4.5 meV which grow in intensity
with Q and therefore correspond mainly to molecular vibrations. Below these frequencies, the
longest simulations (27 ps) indicate that there are no lower lying lattice vibrations implying that the lower energy features observed in the experiment are of magnetic origin.
\section{Summary}

The ${\Cudaca}$ system comprises a very interesting magnetic unit - a Cu${}^{2+}$ S=1/2 tetrahedron.
Due to the geometry of the cluster two exchange paths are possible: antiferromagnetic Cu-O-Cu  via the central oxygen atom and ferromagnetic Cu-Cl-Cu via the peripheral chlorine atoms. In the ideal system the AF exchange with J=-1.8 meV dominates.\\
Our DFT calculations show that unpaired electron is only partially localized in the Cu $d$-states, and hybridizes strongly
 with  the oxygen atom in the center of the tetrahedron and is further delocalized around atomic cores within the molecule. Besides the O$^{2-}$ and Cl$^{-}$ ions, which certainly participate in the Cu-Cu superexchange, the N$\equiv$C bond might play an important role of the charge reservoir.
We plan to confirm this picture by polarised neutron spin density mapping and to study the interplay between electron delocalization and magnetic interactions in more detail.\\ 
The presence of strong molecular vibrations emerge from our INS experiments and spin restricted AIMD simulations. Although all atoms participate in the vibrations, it is the organic ligands and the apical Cl$^{-}$ ions of the cluster which vibrate most strongly, while the O$^{2-}$ ion remains pinned in the center of the tetrahedron.  
Therefore the  spin-vibrational Hamiltonian suggested by Jones\cite{Jones} and Polinger et al.\cite{Polinger1, Polinger2}  might be the appropriate  model to describe the ideal ${\Cudaca}$ system.\\
However, the properties of real samples, cannot be that easily attributed to this model. 
Measurements of the bulk properties of single crystals reveal strong anisotropy which dominates
also the bulk properties of polycrystalline samples. We infer that only part of the clusters retain the spin singlet ground state, while another part has magnetic ground state and the ratio between these fractions is sample and thermal history dependent.\\
The anisotropy originates most probably from
defects introduced during the sample synthesis as there is only small energy difference between several molecular configurations. The isostructural order-disorder structural crossover at T$_{C}$=282 K complicates the situation even more, as applied magnetic field and thermal history may affect which molecule configuration freezes-in determining in
this way the magnetic exchange at low temperatures. 
In this sense the high-temperature manifold of molecular configurations governs magnetic exchange at low temperature. \\
It is well possible that molecular vibrations influence  the low temperature magnetic properties as well,
we leave this question open for future investigations.
We suspect that unusual magnetic susceptibility measured for other representatives of the {\Cucomplex} system\cite{Lines, Blake, Dickinson} might have similar origin to ${\Cudaca}$. It is important to note that influence of the structural and vibrational degrees of freedoms on magnetic properties of molecular magnets is recognized in a number of other systems, for example in $\{V_{15}\}$ complex\cite{Chiorescu00},$\{Ni_{4}Mo_{12}\}$ spin tetrahedron\cite{Schnack06}, $\{Cu_{3}\}$ spin triangle\cite{Choi08}.\\
Finally, the complications of the real samples hinder answering the initial question which motivated our research, namely whether the ground state of an isolated tetrahedron with antiferromagnetic exchange
remains degenerate or whether this degeneracy is easily lifted by some small perturbation. Our study reveals that the behaviour of such "simple" systems is quite complex and
is governed by high sensitivity of electronic structure of the system on the precise molecular configuration and this determines the magnetic behaviour of the material.\\

\section{Acknowledgments}
Discussions with Drs. P. Tregenna-Piggott, D. Chernyshov, A. F. Albuquerque and Prof. S. Klonishner are highly acknowledged.
The work was partially performed at SINQ, Paul Scherrer Insitute, Villigen, Switzerland. We thank for technical support on Quantum Design magnetometers of the Universities of Bern and Z\"{u}rich. 


\end{document}